\documentclass{webofc}
\usepackage[varg]{txfonts}   
\woctitle{The Time Machine Factory 2012}
\begin{document}
\title{Discrete mechanics, ``time machines'' and hybrid systems}
%
%

\author{Hans-Thomas Elze\inst{1}\fnsep\thanks{\email{elze@df.unipi.it}} 
}

\institute{Dipartimento di Fisica ``Enrico Fermi'', Universit\`a di Pisa,  
Largo Pontecorvo 3, I-56127 Pisa, Italia 
          }

\abstract{Modifying the {\it discrete mechanics} proposed by T.D. Lee, 
we construct a class of discrete classical Hamiltonian systems, in which time is one of 
the dynamical variables. This includes a toy model of  ``time machines'' which 
can travel forward and backward in time and which differ from models based on 
closed timelike curves (CTCs). In the continuum limit, we explore the 
interaction between such time reversing machines and quantum mechanical objects, 
employing a recent description of quantum-classical hybrids. 

}
\maketitle 

\section{Introduction}
\label{Intro}
Time travel and time machines have been the stuff of science fiction for a while and possibly  excited human minds much earlier than that. -- However, they have become 
a topic of active scientific enquiry since the realization that 
certain cosmological solutions of Einstein's equations of general relativity  allow 
for {\it closed timelike curves} (CTCs). Here an object can travel in an unusual 
geometry of spacetime, such that it encounters the past and, in particular, its own 
past. It is obvious that - with the link between quantum mechanics and 
general relativity still little understood - this provides an arena for producing 
paradoxes ({\it e.g.}, grandfather paradox, unproved theorem paradox) and testing new 
ideas how to resolve them, besides availing surprising computational resources. 
-- 
Discussion of  background, an overview of  
existing literature, and the state of the art of constructing {\it quantum mechanical time machines} can be found in Refs.\, \cite{Lloydetal11a,Lloydetal11b}. 
 
Our aim here is threefold. -- In Sect.~\ref{DiscrHamil}, we recall T.D. Lee's 
proposal of {\it time as a fundamentally discrete dynamical  variable} \cite{Lee83,Lee87}. 
Limiting the number of events or measurements in a given spacetime region, this can  
have surprising consequences in the continuum limit. 
We modify his action principle in such a way that a 
Hamiltonian formulation for such discrete systems becomes available. 
Furthermore, we show that these systems allow for a particular kind of time 
machines, namely {\it time reversing machines}. 
-- 
In Sect.~\ref{QuClHybr}, we review a recent attempt to construct a theory that 
describes {\it quantum-classical hybrids}, consisting of quantum mechanical and 
classical objects that interact directly with each other \cite{me11,me12,me12b}. 
Hybrids might exist as a fundamentally different species of composite objects 
"out there", 
with consequences for the range of applicability of quantum mechanics, or they may 
serve as approximate description for certain complex quantum systems. 

We employ the concept of quantum-classical  hybrids, in order to 
explore a hypothetical {\it direct coupling of classical time machines to quantum 
objects}. In Sect.~\ref{QuContr}, we introduce a specific model, obtain its 
equations of motion, and discuss consequences for the study of time machines, 
followed by conclusions in Sect.~\ref{Conclu}. 

\newpage
\section{Discrete Hamiltonian mechanics}
\label{DiscrHamil}  
Discrete dynamical systems arise in many contexts in physics or mathematics, for 
example, in discrete approximations or maps facilitating numerical studies of complex systems, as regularized versions of quantum field theories on spacetime lattices, or 
describing intrinsically discrete processes.  Here, the usual preponderance of differential  
equations over finite difference equations is given up for reasons which may be 
more or less fundamental.   

In a series of articles T.D. Lee and collaborators have proposed to incorporate discreteness as a fundamental aspect of dynamics, see Refs.\,\cite{Lee83,Lee87} and 
further references therein, and have elaborated various classical and quantum 
models in this vein, which share desirable symmetries with the corresponding 
continuum theories  while presenting finite degrees of freedom.
  
For our purposes, it will be sufficient to consider classical discrete mechanics which derives from the basic assumption that {\it time is a discrete dynamical variable}. This 
naturally invokes a {\it fundamental length or time} (in natural units), $l$. Which can be rephrased as the assumption that in a fixed $(d+1)$-dimensional spacetime volume $\Omega$ maximally $N=\Omega /l^{d+1}$ measurements can be performed or this number of events take place. 
 
In Refs.\,\cite{Lee83,Lee87} a variational principle was presented, based on a 
Lagrangian formulation of the postulated action. Various forms and (dis)advantages 
of such an approach have subsequently been discussed, {\it e.g.}, in 
Refs.\,\cite{DInnocenzo87,HongBin05,Toffoli11}. Presently, we present a Hamiltonian 
formulation which differs from all previous ones in that it leads to particularly 
transparent and symmetric equations of motion. This allows us to introduce a 
suitable Poisson bracket and a phase space description of the dynamics. The latter 
is an essential ingredient when constructing quantum-classical hybrids, 
as we shall see in the following Sect.~\ref{QuClHybr}. 
  
We describe the state $n$ of a discrete mechanical object by its positions in 
spacetime in terms of the real {\it dynamical variables} $x_n,\tau_n$ and corresponding 
{\it conjugated momenta} $p_n,{\cal P}_n$, with $n=0,1,2,\dots$\,.\footnote{$x_n$ and $p_n$ might be vectors, depending on the dimensionality of space, while $\tau_n$ and ${\cal P}_n$ are assumed onedimensional.} 
We postulate that its dynamics is governed by the stationarity of this {\it action}: 
\begin{equation}\label{action} 
A:=
\sum_{n>0}\Big [(p_n+p_{n-1})\Delta x_n+({\cal P}_n+{\cal P}_{n-1})\Delta\tau_n
-{\cal H}_n\Big ]  
\;\;, \end{equation} 
under independent variations of all variables and momenta; 
the finite differences are defined by: 
\begin{equation}\label{findiff}
\Delta x_n:=x_n-x_{n-1}\;\;,\;\;\;\Delta\tau_n:=\tau_n-\tau_{n-1}
\;\;, \end{equation} 
and the Hamiltonian function by ${\cal H}:=\sum_n{\cal H}_n$, with:  
\begin{equation}\label{Hamiltonian} 
{\cal H}_n:=\Delta\tau_n\Big [\frac{p_n^{\;2}+p_{n-1}^{2}}{2}+V(x_n)+V(x_{n-1})\Big ] 
+{\cal K}_n 
\;\;, \end{equation} 
where $V$ is a sufficiently smooth potential and ${\cal K}_n$ will be specified  
in due course. 

The variations of the action amount to differentiations here and lead to the   
equations of motion: 
\begin{eqnarray}\label{eom1}
\dot x_n&=&\dot\tau_np_n+\partial_{p_n}\sum_{n'}{\cal K}_{n'} 
\;=\;\partial_{p_n}{\cal H}\;\equiv\;\{x_n,{\cal H}\}
\;\;, \\ [1ex]\label{eom2} 
\dot p_n&=&-\dot\tau_n\partial_{x_n}V(x_n)
-\partial_{x_n}\sum_{n'}{\cal K}_{n'}\;=\;-\partial_{x_n}{\cal H}\;\equiv\;\{p_n,{\cal H}\}
\;\;, \\ [1ex]\label{eom3} 
\dot\tau_n&=&\partial_{{\cal P}_n}\sum_{n'}{\cal K}_{n'}
\;=\;\partial_{{\cal P}_n}{\cal H}\;\equiv\;\{\tau_n,{\cal H}\}
\;\;, \\ [1ex]\label{eom4} 
\dot{\cal P}_n&=&E_{n+1}-E_n-\partial_{\tau_n}\sum_{n'}{\cal K}_{n'}
\;\equiv\;\{{\cal P}_n,{\cal H}\}
\;\;, \end{eqnarray}
introducing the discrete ``time derivative'',  $\dot O_n:=O_{n+1}-O_{n-1}\;$, on the 
left-hand and Poisson brackets (cf. below) on the right-hand sides, respectively; 
furthermore, $E_n:=\frac{1}{2}(p_n^{\;2}+p_{n-1}^{\;2})+V(x_n)+V(x_{n-1})\;$.

Several remarks are in order here. -- Assuming that ${\cal K}_n$ depends only on 
the state $n$, it can be shown that the set of Eqs.\,(\ref{eom1})--(\ref{eom4}) is 
{\it time  reversal invariant}; the state $n+1$ can be calculated from knowledge 
of the earlier states $n$ and $n-1$ and the state $n-1$ from the later ones 
$n+1$ and $n$.   

Furthermore, stationarity of the action under independent variations of $\tau_n$ and 
$x_n$ for every state $n$ implies 
{\it invariance under translations in  time and space}, respectively,  
and the conservation of energy and momentum (modulo the effect of the external 
force deriving from $V$). This holds under the further assumption that ${\cal K}_n$ does 
{\it not} enter through its derivatives in Eqs.\,(\ref{eom1})--(\ref{eom2}). 

Explicit solutions of the equations of motion can be easily found 
in the case that ${\cal K}_n:=0$ and the potential $V$ is constant or a linear function, 
{\it i.e.}, for zero or constant external force . This recovers the behaviour discussed in 
Refs.\,\cite{Lee83,DInnocenzo87}. However, in the present formulation, we also 
have the possibility to study more exotic models, in which the dynamics of 
the time variable $\tau_n$ itself plays an important role when ${\cal K}_n\neq0$, 
cf. Sect.~\ref{2.2}.     

Considering  the dynamical variables and canonically conjugated momenta as 
canonical coordinates for the phase space spanned by 
$\{ x_n,\tau_n;p_n,{\cal P}_n\}$,  we 
introduce the {\it Poisson bracket} of any two regular functions $f$ 
and $g$ on this space: 
\begin{equation}\label{discPoisson} 
\{ f,g\}:=\sum_n\Big (\frac{\partial f}{\partial_{x_n}}\frac{\partial g}{\partial_{p_n}}
-\frac{\partial f}{\partial_{p_n}}\frac{\partial g}{\partial_{x_n}}
+\frac{\partial f}{\partial_{\tau_n}}\frac{\partial g}{\partial_{{\cal P}_n}}
-\frac{\partial f}{\partial_{{\cal P}_n}}\frac{\partial g}{\partial_{\tau_n}}\Big ) 
\;\;, \end{equation} 
which has been indicated already on the right-hand sides of 
Eqs.\,(\ref{eom1})--(\ref{eom2}),     

This allows a convenient description also of ensembles of discrete mechanical 
objects, which individually follow the above equations of motion. Let us 
collectively denote variables and momenta as $Q_n$ and $P_n$, respectively, such that   
$\{ f,g\}=\sum_n\Big (\frac{\partial f}{\partial_{Q_n}}\frac{\partial g}{\partial_{P_n}}
-\frac{\partial f}{\partial_{P_n}}\frac{\partial g}{\partial_{Q_n}}\Big )$. Then, we postulate, 
in analogy to continuum mechanics (see Subsect.~\ref{2.1}), a {\it continuity equation}  
to determine the flow of the probability density $\rho_n\equiv\rho_n(Q_n;P_n)$ of the 
ensemble in phase space:  
\begin{equation}\label{continuityeq} 
0=\dot\rho_n+\partial_{Q_n}(\rho_n\dot Q_n)
+\partial_{P_n}(\rho_n\dot P_n) 
\;\;, \end{equation} 
with $\dot O_n:=O_{n+1}-O_{n-1}\;$, as before. Employing the equations of 
motion (\ref{eom1})--(\ref{eom4}), this continuity equation can be rewritten as 
the discrete mechanics analogue of the {\it Liouville equation}: 
\begin{equation}\label{discLiouville} 
\dot\rho_n=\{ {\cal H},\rho_n\} 
\;\;. \end{equation} 
In the following, we study the continuum limit of the equations of motion, in which time 
remains one of the dynamical variables. 

\subsection{The continuum limit}
\label{2.1}
In order to discuss the continuum limit, we let the fundamental time (or length) 
constant  become arbitrarily small, $l\rightarrow 0$, such that the density of events 
or measurements becomes correspondingly large, $N\rightarrow\infty$.   
Furthermore, we introduce the {\it external time}, $t:=nl$, with $n=0,1,2,\dots$\,, and 
define $x(t):=x_n$, $\tau (t):=\tau_n$, $p(t):=p_n$, ${\cal P}(t):={\cal P}_n$, {\it i.e.}, in terms 
of the discrete dynamical variables and conjugated momenta. Thus, for example, 
$\tau_{n+1}-\tau_n=\tau (t+l)-\tau (t)=\dot\tau (t)l+\mbox{O}(l^2)\;$, where   
$\dot\tau :=\mbox{d}\tau /\mbox{d}t$, {\it etc.} -- 
In this way, we obtain the equations of motion in the continuum limit: 
\begin{eqnarray}\label{eom1c}
\dot x&=&\dot\tau p+\frac{1}{2l}\partial_{p_n}\sum_{n'}{\cal K}_{n'} 
\;\;, \\ [1ex]\label{eom2c} 
\dot p&=&-\dot\tau\nabla V(x)
-\frac{1}{2l}\partial_{x_n}\sum_{n'}{\cal K}_{n'} 
\;\;, \\ [1ex]\label{eom3c} 
\dot\tau&=&\frac{1}{2l}\partial_{{\cal P}_n}\sum_{n'}{\cal K}_{n'}
\;\;, \\ [1ex]\label{eom4c} 
\dot{\cal P}&=&\frac{\mbox{d}}{\mbox{d}t}\Big [\frac{p^2}{2}+V(x)\Big ]
-\frac{1}{2l}\partial_{\tau_n}\sum_{n'}{\cal K}_{n'}
\;\;, \end{eqnarray} 
where terms containing $\sum {\cal K}_{n'}$ will be defined and evaluated shortly.   

It suffices here to assume that all ${\cal K}_{n'}$ are independent of $\{x_n,p_n\}$.  
This simplifies  Eqs.\,(\ref{eom1c})--(\ref{eom2c}): 
\begin{equation}\label{eom12c} 
\dot x=\dot\tau p \;\;,\;\;\; 
\dot p=-\dot\tau\nabla V(x)
\;\;, \end{equation} 
implies $\mbox{d}/\mbox{d}t[p^2/2+V(x)]=0\;$, and, consequently, simplifies 
also Eq.\,(\ref{eom4c}). In this case, $\dot\tau$ plays the role of a given 
{\it ``lapse'' function} for the subsystem described by $x$ and $p$, which  
can be separately determined  
(cf. Sect.\,2.2). {\i I.e.},  if Eqs.\,(\ref{eom3c})--(\ref{eom4c}) are integrated explicitly, the remaining Eqs.\,(\ref{eom12c}) follow from the time dependent effective Hamiltonian: 
\begin{equation}\label{effH} 
{\cal H}_c(x,p;t):=\dot\tau (t) [\frac{p^2}{2}+V(x)]  
\;\;, \end{equation} 
with $\dot\tau$ as a time dependent parameter.

The existence of a simple continuum Hamiltonian, such as ${\cal H}_c$, is not  
obvious, in general, since  $\Delta\tau_n$ on the 
right-hand side of Eq.\,(\ref{Hamiltonian}) becomes proportional to $\dot\tau$, if one  
performs the continuum limit directly on the discrete dynamics 
Hamiltonian; the presence of this factor can spoil the Hamiltonian picture of the 
resulting dynamics.  

\subsection{Time machines}
\label{2.2} 
Here we illustrate the continuum 
limit of the discrete mechanics that we obtained. We choose 
$ {\cal K}_n:=l[{\cal P}_n^{\;2}+{\cal V}(\tau_n)]\;$. Then, the continuum limit 
applied to Eqs.\,(\ref{eom3c})--(\ref{eom4c}) gives simply: 
\begin{equation}\label{eom34c}   
\dot\tau ={\cal P}\;\;,\;\;\;\dot{\cal P}=
-\frac{1}{2}\frac{\mbox{d}}{\mbox{d}\tau}{\cal V}(\tau) 
\;\;. \end{equation}  
with $\dot\tau :=\mbox{d}\tau /\mbox{d}t$, {\it etc.} 

We observe that for suitable potentials ${\cal V}(\tau)$ and initial conditions 
the {\it internal time} $\tau$ will perform a {\it bounded periodic motion} as function 
of the {\it external time} $t$. For example, for an oscillator potential, 
${\cal V}(\tau):=\omega^2\tau^2$, we obtain solutions 
$\tau (t)=\bar\tau\sin (\omega t)$, with amplitude $\bar\tau$ and phase determined 
by the initial conditions, such that $\dot\tau (t)=\dot\tau (-t)$ is time reversal 
invariant.  

Furthermore, the Eqs.\,(\ref{eom12c}) can be rewritten as a single second 
order equation: 
\begin{equation}\label{eom12c2} 
\frac{\mbox{d}^2}{\mbox{d}\tau^2}x=-\nabla V(x)
\;\;,\;\;\mbox{with}\;\tau\equiv\tau (t) 
\;\;, \end{equation} 
{\it i.e.}, as an ordinary equation of motion with respect to the internal time, 
which is considered as a function of the external time, to be obtained from 
Eqs.\,(\ref{eom34c}).   

This situation describes a toy model of {\it time machines}:  
the $x,p$-subsystem moves forward in time on a particular trajectory in 
phase space, as long as $\tau (t)$ increases; when, due to its periodicity, this 
function decreases, this trajectory is traced identically backwards!  Thus,  
the behaviour in the external time $t$ is cyclic, alternating between forward and 
backward evolution.   

We remark that this dynamical implementation of ``time travel'' differs 
from a frequently considered one, which is based on modifying the background 
spacetime structure. In particular, Politzer's spacetime, which allows 
closed timelike curves (CTCs),   
is obtained by identifying a certain spatial region at one time with the same region at a 
later time \cite{Politzer92}; thus, an object may {\it transit} instantaneously from a final 
state to the corresponding (identical) initial state of its evolution. In our model, it {\it evolves} identically backwards  from a final state to its initial state; it is conceivable that 
this can be realized in physical analogue models. 

In Sect.~\ref{QuContr}, we explore the coupling of such a classical time machine to a quantum object in a particular framework describing quantum-classical hybrids. 

\section{Quantum-classical hybrids}
\label{QuClHybr}  
The direct coupling of quantum mechanical (QM) and classical (CL) degrees of 
freedom -- {\it ``hybrid dynamics''} -- departs from quantum mechanics. We summarize   
here briefly the theory presented in Refs.\,\cite{me11,me12,me12b}, where also 
additional  references and discussion of related works can be found.  

Hybrid dynamics has been researched extensively for various reasons. -- 
For example, the Copenhagen interpretation of quantum mechanics entails the  
measurement problem which, together with the fact that quantum mechanics needs  
interpretation, in order to be operationally well defined, may indicate that 
it needs amendments. Such as   
a theory of the {\it dynamical} coexistence of  QM and CL objects.   
This should have  
impact on the measurement problem \cite{Sudarshan123} as well as on  
the description of the interaction between quantum matter and (possibly) classical 
spacetime \cite{BoucherTraschen}.  

Furthermore, it is of great practical interest to better 
understand QM-CL hybrids appearing in QM approximation 
schemes addressing many-body systems or interacting fields, 
which are naturally separable into QM and CL subsystems; for example, representing 
fast and slow degrees of freedom, mean fields and fluctuations, {\it etc.}  
 
Concerning the hypothetical emergence of quantum mechanics from some  
coarse-grained deterministic dynamics (see Refs.\,\cite{tHooft10,Elze09a,Adler} 
with numerous references to related work), the quantum-classical backreaction 
problem might  appear in new form, namely regarding the interplay of fluctuations 
among underlying deterministic 
and emergent QM degrees of freedom. Which can be 
rephrased succinctly as: {\it ``Can quantum mechanics be seeded?''}

Thus, there is ample motivation for the numerous attempts to formulate a satisfactory 
 hybrid dynamics. Generally, they are deficient in one or another respect. 
Which has led to various no-go theorems, in view of the lengthy list of desirable properties or consistency requirements that ``{\it the}'' hybrid theory should 
fulfil, see, for example, 
Refs.\,\cite{CaroSalcedo99,DiosiGisinStrunz}:  
\begin{itemize} 
\item Conservation of energy. 
\item Conservation and positivity of probability. 
\item Separability of QM and CL subsystems in the absence of their interaction, 
recovering the correct QM and CL equations of motion, respectively. 
\item Consistent definitions of states and observables; existence of a Lie bracket structure 
on the algebra of observables that suitably generalizes  
Poisson and commutator brackets. 
\item Existence of canonical transformations generated by the observables; 
invariance of the classical sector under canonical transformations 
performed on the quantum sector only and {\it vice versa}. 
\item Existence of generalized Ehrenfest relations ({\it i.e.} the 
correspondence limit) which, 
for bilinearly coupled CL and QM oscillators,  
are to assume the form of the CL equations of motion  
(``Peres-Terno benchmark'' test \cite{PeresTerno}). 
\item `Free Will' \cite{Diosi11}. 
\item Locality. 
\item No-signalling. 
\item QM / CL symmetries and ensuing separability carry over to hybrids. 
\end{itemize} 

These issues have also been discussed for the hybrid ensemble theory of Hall and
Reginatto, which  does conform with the first six points listed  but is in conflict with 
the last two \cite{HallReginatto05,Hall08}.  

We have proposed an alternative theory 
of hybrid dynamics based on notions of phase space \cite{me11}. 
This extends work by Heslot, demonstrating   
that quantum mechanics can entirely be rephrased in the language and 
formalism of classical analytical mechanics \cite{Heslot85}. 
Introducing unified notions 
of states on phase space, observables, canonical transformations, and a generalized 
quantum-classical Poisson bracket, this has led to an intrinsically 
linear hybrid theory, which allows to fulfil {\it all} of the above consistency requirements. 

Recently Buri\'c and collaborators have shown that the dynamical aspects of our 
proposal can indeed be derived for an all-quantum mechanical composite system by 
imposing constraints on fluctuations in one subsystem, followed by suitable coarse-graining \cite{Buric12,Buric12b}. 

Besides constructing the QM-CL hybrid formalism and showing how 
it conforms with the above consistency requirements, we earlier discussed  
the possibility to have classical-environment induced decoherence, quantum-classical 
backreaction, a deviation from the Hall-Reginatto proposal in presence 
of translation symmetry, and closure of the algebra of hybrid observables 
\cite{me11,me12b}. Questions of locality, symmetry vs. separability, incorporation of 
 superposition, separable, and entangled QM states, and `Free Will' were considered 
in Ref.\,\cite{me12}. 

\subsection{Quantum mechanics -- rewritten in classical terms}
\label{QuMech} 
We recall that evolution of a {\it classical} object can be described in relation to its 
$2n$-dimensional phase space, its {\it state space}. A real-valued regular 
function on this space defines an {\it observable}, {\it i.e.}, a differentiable function 
on this smooth manifold.  

There always exist (local) systems of  
{\it canonical coordinates}, commonly denoted by $(x_k,p_k),\; k=1,\dots ,n$, 
such that the {\it Poisson bracket} of any pair of observables $f,g$ assumes 
the standard form: 
\begin{equation}\label{PoissonBracket} 
\{ f,g \}\; =\; 
\sum_k\Big (\frac{\partial f}{\partial_{x_k}}\frac{\partial g}{\partial_{p_k}}
-\frac{\partial f}{\partial_{p_k}}\frac{\partial g}{\partial_{x_k}}\Big ) 
\;\;. \end{equation} 
This is consistent with $\{ x_k,p_l\}=\delta_{kl}$, $\{ x_k,x_l\} =\{ p_k,p_l\}=0,\; 
k,l=1,\dots ,n$, and has the properties defining a Lie bracket operation, 
mapping a pair of observables to an observable.  

General transformations ${\cal G}$ of the state space are restricted by compatibility with 
the Poisson bracket structure to so-called {\it canonical transformations}, which   
do not change physical properties of an object. They 
form a Lie group and it is sufficient 
to consider infinitesimal transformations. 
An {\it infinitesimal transformation} ${\cal G}$ is 
{\it canonical}, if and only if for any observable $f$ the map $f\rightarrow {\cal G}(f)$ 
is given by $f\rightarrow f'=f+\{ f,g\}\delta\alpha$, with some observable $g$, 
the so-called {\it generator} of ${\cal G}$, and $\delta\alpha$ an infinitesimal real 
parameter. -- Thus, for example, the canonical coordinates transform as follows: 
\begin{equation}\label{xpcan} 
x_k\;\rightarrow\;x_k'=x_k+\frac{\partial g}{\partial p_k}\delta\alpha 
\;\;,\;\;\;   
p_k\;\rightarrow\;p_k'=p_k-\frac{\partial g}{\partial x_k}\delta\alpha 
\;\;. \end{equation} 
This illustrates the fundamental relation between observables and generators 
of infinitesimal canonical transformations in classical Hamiltonian mechanics. 
\vskip 0.5cm 

Following Heslot's work, we learn that the previous analysis 
can be generalized and applied to quantum mechanics; this concerns 
the dynamical aspects as well as the notions of states, canonical transformations, and 
observables \cite{Heslot85}. 
\vskip 0.5cm 

The {\it Schr\"odinger equation} and its adjoint can be 
derived as Hamiltonian equations from an action principle 
\cite{me11}. We must add the {\it normalization condition}, 
${\cal C}:=\langle\Psi (t)|\Psi (t)\rangle\stackrel{!}{=}1\;$, 
for all state vectors $|\Psi\rangle$, which is    
essential for the probability interpretation of amplitudes;    
state vectors that differ by an unphysical constant phase are to be 
identified. Thus, the {\it QM state space} 
is formed by the rays of the underlying Hilbert space. 

\subsubsection{Oscillator representation}
\label{OscRepr}
A unitary transformation describes QM evolution, 
$|\Psi (t)\rangle =\hat U(t-t_0)|\Psi (t_0)\rangle$, 
with $U(t-t_0)=\exp [-i\hat H(t-t_0)/\hbar ]$, 
solving the Schr\"odinger equation. Thus, a stationary state, characterized by 
$\hat H|\phi_i\rangle =E_i|\phi_i\rangle$, with real energy eigenvalue $E_i$,  
performs a harmonic motion, {\it i.e.}, 
$|\psi_i(t)\rangle =\exp [-iE_i(t-t_0)/\hbar ]|\psi_i(t_0)\rangle
\equiv\exp [-iE_i(t-t_0)/\hbar ]|\phi_i\rangle$. We assume a denumerable set of 
such states. Following these observations, it is quite natural to introduce 
the following {\it oscillator representation}. 

We expand state vectors with respect to a complete 
orthonormal basis, $\{ |\Phi_i\rangle\}$:  
\begin{equation}\label{oscillexp} 
|\Psi\rangle =\sum_i|\Phi_i\rangle (X_i+iP_i)/\sqrt{2\hbar} 
\;\;, \end{equation} 
where the time dependent coefficients are separated into  
real and imaginary parts, $X_i,P_i$ \cite{Heslot85}. 
This expansion allows to   
evaluate what we {\it define} as {\it Hamiltonian function}, {\it i.e.},   
${\cal H}:=\langle\Psi |\hat H|\Psi\rangle$: 
\begin{equation}\label{HamiltonianQM1} 
{\cal H}=\frac{1}{2\hbar}\sum_{i,j}\langle\Phi_i|\hat H|\Phi_j\rangle (X_i-iP_i)(X_j+iP_j) 
=:{\cal H}(X_i,P_i) 
\;\;. \end{equation} 
Choosing the set of energy eigenstates, $\{ |\phi_i\rangle\}$, 
as basis of the expansion, we obtain:     
\begin{equation}\label{HamiltonianQM2} 
{\cal H}(X_i,P_i)=\sum_i\frac{E_i}{2\hbar}(P_i^{\;2}+X_i^{\;2}) 
\;\;, \end{equation} 
hence the name {\it oscillator representation}. --  
Evaluating $|\dot\Psi\rangle =
\sum_i|\Phi_i\rangle (\dot X_i+i\dot P_i)/\sqrt{2\hbar}$ 
according to Hamilton's equations with ${\cal H}$ of 
Eq.\,(\ref{HamiltonianQM1}) or (\ref{HamiltonianQM2}), gives back the Schr\"odinger equation. -- 
Furthermore, the {\it normalization condition} becomes: 
\begin{equation}\label{oscillnormalization} 
{\cal C}(X_i,P_i)=\frac{1}{2\hbar}\sum_i(X_i^{\;2}+P_i^{\;2})\stackrel{!}{=}1 
\;\;. \end{equation} 
Thus, the vector with components given by 
$(X_i,P_i),\; i=1,\dots ,N$, is 
confined to the surface of a $2N$-dimensional sphere with radius $\sqrt{2\hbar}$, 
which presents a major difference to CL Hamiltonian mechanics. 

The $(X_i,P_i)$ may be considered 
as {\it canonical coordinates} for the state space of a 
QM object. Correspondingly, we introduce a {\it Poisson bracket}, 
cf. Eq.(\ref{PoissonBracket}), 
for any two {\it observables} on the {\it spherically compactified state space}, 
{\it i.e.} real-valued regular functions $F,G$ of the coordinates $(X_i,P_i)$:     
\begin{equation}\label{QMPoissonBracket} 
\{ F,G \}\; =\; 
\sum_i\Big (\frac{\partial F}{\partial_{X_i}}\frac{\partial G}{\partial_{P_i}}
-\frac{\partial F}{\partial_{P_i}}\frac{\partial G}{\partial_{X_i}}\Big ) 
\;\;. \end{equation} 
As usual, time evolution of an observable $O$ is generated by the Hamiltonian: 
$\mbox{d}O/\mbox{d}t=\partial_tO+\{ O,{\cal H} \}\;$.  
In particular, we find that the constraint of Eq.\,(\ref{oscillnormalization}) 
is conserved:  $\mbox{d}{\cal C}/\mbox{d}t=\{ {\cal C},{\cal H} \}=0\;$.

\subsubsection{Canonical transformations and quantum observables} 
\label{CanTransObserv}
In the following, we recall briefly the compatibility of the notion of 
observable introduced in passing above -- as in classical mechanics -- with the usual 
QM one. This can be demonstrated rigourously by the implementation of canonical transformations  and analysis of the role of observables as their generators.  
For details, see Refs.\,\cite{me11,me12,me12b,Heslot85}.   

The Hamiltonian function has been defined as observable in  
Eq.\,(\ref{HamiltonianQM1}), which relates it directly to the corresponding QM   
observable, namely the expectation of the self-adjoint Hamilton operator. 
This is indicative of the general structure with the following most important 
features:  
\\ \noindent $\bullet$ 
A) {\it Compatibility of unitary transformations and Poisson structure.} -- 
Classical canonical transformations are automorphisms of the state space 
which are compatible with the Poisson bracket. 
Automorphisms of the QM Hilbert space are implemented by 
unitary transformations. This implies a transformation of the canonical 
coordinates $(X_i,P_i)$ here. From this, one derives the invariance of   
the Poisson bracket defined in Eq.\,(\ref{QMPoissonBracket}) under unitary transformations. 
Consequently, the {\it unitary transformations on Hilbert space are canonical transformations on the $(X,P)$ state space}.   
\\ \noindent $\bullet$ 
B) {\it Self-adjoint operators as observables.} -- 
Any infinitesimal unitary transformation $\hat U$ can be generated by a self-adjoint operator 
$\hat G$, such that: $\hat U=1-(i/\hbar )\hat G\delta\alpha\;$,  
which leads to the QM relation between an observable and 
a self-adjoint operator. By a simple calculation, one obtains: 
\begin{equation}\label{XPcan} 
X_i\;\rightarrow\; X_i'=X_i+\frac{\partial \langle\Psi |\hat G|\Psi\rangle}
{\partial P_i}\delta\alpha 
\;\;,\;\;\; 
P_i\;\rightarrow\; P_i'=P_i-\frac{\partial \langle\Psi |\hat G|\Psi\rangle}
{\partial X_i}\delta\alpha 
\;\;. \end{equation} 
From these equations, the  
relation between an observable $G$, defined in analogy to classical mechanics 
(as above), and a self-adjoint 
operator $\hat G$ can be inferred: 
\begin{equation}\label{goperator} 
G(X_i,P_i)=\langle\Psi |\hat G|\Psi\rangle 
\;\;, \end{equation} 
{\it i.e.}, by comparison with the classical result.  
Hence, a {\it real-valued regular function $G$ of the state is an observable, if 
and only if there exists a self-adjoint operator $\hat G$ such that Eq.\,(\ref{goperator}) 
holds}. This implies that {\it all QM observables are quadratic forms} 
in the $X_i$'s and $P_i$'s, which are essentially fewer than in the corresponding CL 
case; interacting QM-CL hybrids require additional discussion, see Ref.\,\cite{me12b}.  
\\ \noindent $\bullet$ 
C) {\it Commutators as Poisson brackets.} -- 
From the relation (\ref{goperator}) between observables and self-adjoint operators and the  Poisson bracket (\ref{QMPoissonBracket}) one derives: 
\begin{equation}\label{QMPBComm} 
\{ F,G\}=\langle\Psi |\frac{1}{i\hbar}[\hat F,\hat G]|\Psi\rangle 
\;\;, \end{equation}  
with both sides of the equality considered as functions of the variables $X_i,P_i$ 
and with the commutator defined as usual. Hence,  
the {\it QM commutator is a Poisson bracket with respect to the $(X,P)$ state space} and 
relates the algebra of its observables to the algebra of self-adjoint operators.   
\vskip 0.5cm

In conclusion, quantum mechanics shares with classical mechanics an even dimensional 
state space, a Poisson structure, and a related algebra of observables. It  
differs essentially by a restricted set of observables and the requirements 
of phase invariance and normalization, which compactify the underlying Hilbert space 
to the complex projective space formed by its rays.  

\subsection{Quantum-classical Poisson bracket, hybrid states and their evolution}
\label{PoissonBrEvol}
The far-reaching parallel of classical and quantum mechanics, as we have seen,    
suggests to introduce a {\it generalized Poisson bracket} for QM-CL hybrids:  
\begin{eqnarray}\label{GenPoissonBracket} 
\{ A,B\}_\times &:=&\{ A,B\}_{\mbox{\scriptsize CL}}+\{ A,B\}_{\mbox{\scriptsize QM}}
\\ [1ex] \label{GenPoissonBracketdef} 
&:=&\sum_k\Big (\frac{\partial A}{\partial_{x_k}}\frac{\partial B}{\partial_{p_k}}
-\frac{\partial A}{\partial_{p_k}}\frac{\partial B}{\partial_{x_k}}\Big )+  
\sum_i\Big (\frac{\partial A}{\partial_{X_i}}\frac{\partial B}{\partial_{P_i}}
-\frac{\partial A}{\partial_{P_i}}\frac{\partial B}{\partial_{X_i}}\Big ) 
\;\;, \end{eqnarray} 
of any two observables $A,B$ defined on the Cartesian product of CL {\it and} QM 
state spaces. It shares the usual properties of a Poisson bracket. -- Note that due to the 
convention  introduced by Heslot \cite{Heslot85}, to which we adhered in 
Sect.~\ref{QuMech},  the QM variables $X_i,P_i$ have dimensions of 
(action$)^{1/2}$ and, consequently, no $\hbar$ appears in 
Eqs.\,(\ref{GenPoissonBracket})--(\ref{GenPoissonBracketdef}). At the expense 
of introducing appropriate rescalings, these variables could be made to have their 
usual dimensions and $\hbar$ to appear explicitly here. -- For the remainder  
of this article, instead we choose units such that $\hbar\equiv 1$.
 
Let an {\it observable ``belong'' to the CL (QM) sector, if it is  
constant with respect to the canonical coordinates of the QM (CL) sector}. Then,   
the $\{\;,\;\}_\times$-bracket has the important properties: 
\\ \noindent $\bullet$ 
D) It reduces to the Poisson brackets introduced 
in Eqs.\,(\ref{PoissonBracket}) and (\ref{QMPoissonBracket}), respectively,   
for pairs of observables that belong {\it either} to the CL {\it or} the QM sector. 
$\bullet$ 
E) It reduces to the appropriate one of the former brackets, 
if one of the observables belongs only to either one of the two sectors. 
$\bullet$ 
F) It reflects the {\it separability} of CL and QM sectors, 
since $\{ A,B\}_\times =0$, if $A$ and $B$ belong to different sectors.  
Hence, {\it if a canonical tranformation 
is performed on the QM (CL) sector only, then observables that belong to the 
CL (QM) sector remain invariant.}

The hybrid density $\rho$ for a self-adjoint density operator $\hat\rho$ in a given 
state $|\Psi\rangle$ is defined by  \cite{me11} :  
\begin{equation}\label{rhodens}
\rho (x_k,p_k;X_i,P_i):=\langle\Psi |\hat\rho (x_k,p_k)|\Psi\rangle
=\frac{1}{2}\sum_{i,j}\rho_{ij}(x_k,p_k)(X_i-iP_i)(X_j+iP_j)
\;\;, \end{equation} 
using  Eq.\,(\ref{oscillexp}) and   
$\rho_{ij}(x_k,p_k):=\langle\Phi_i|\hat\rho (x_k,p_k)|\Phi_j\rangle 
=\rho_{ji}^\ast (x_k,p_k)$. It describes a {\it QM-CL hybrid ensemble} 
by a real-valued, positive semi-definite, normalized, and possibly time dependent 
regular function on the Cartesian 
product state space canonically coordinated by $2(n+N)$-tuples $(x_k,p_k;X_i,P_i)$; 
the variables $x_k,p_k,\; k=1,\dots ,n$ and $X_i,P_i,\; i=1,\dots ,N$ are reserved 
for CL and QM sectors, respectively. 

It can be shown 
that $\rho (x_k,p_k;X_i,P_i)$ is the {\it probability density to find in the hybrid 
ensemble the QM state} $|\Psi\rangle$, parametrized by $X_i,P_i$ through 
Eq.\,(\ref{oscillexp}), {\it together with  the CL state} given by a point in phase space, 
specified by coordinates $x_k,p_k$.  -- Further remarks,  
concerning superposition, pure/mixed, or separable/entangled QM states that may enter
the hybrid density can be found in Ref.\,\cite{me12}.  

Furthermore, the simple form of $\rho$ as bilinear function of QM 
``phase space'' variables $X_i,P_i$, stemming from the expectation of a density 
operator $\hat\rho$, has to be generalized for interacting hybrids,  
allowing for so-called {\it almost-classical observables}; see Sect.~5.4 of 
Ref.\,\cite{me11} and a related study \cite{me12b}. 

We are now in the position to introduce the appropriate {\it Liouville equation} 
for the dynamical evolution of hybrid ensembles \cite{me11}. 
Based on Liouville's theorem and the generalized Poisson bracket defined in 
Eqs.\,(\ref{GenPoissonBracket})--(\ref{GenPoissonBracketdef}), we are led to: 
\begin{equation}\label{rhoevol} 
-\partial_t\rho = \{\rho ,{\cal H}_\Sigma\}_\times  
\;\;, \end{equation}
with ${\cal H}_\Sigma\equiv{\cal H}_\Sigma (x_k,p_k;X_i,P_i)$ and:  
\begin{equation}\label{HtotalInt} 
{\cal H}_\Sigma:={\cal H}_{\mbox{\scriptsize CL}}(x_k,p_k)
+{\cal H}_{\mbox{\scriptsize QM}}(X_i,P_i) 
+{\cal I}(x_k,p_k;X_i,P_i)   
\;\;, \end{equation} 
which defines the relevant Hamiltonian function, including a hybrid interaction;  
${\cal H}_\Sigma$ is required to be an {\it observable}, in order to have 
a meaningful notion of energy. Note that {\it energy conservation} follows from 
$\{{\cal H}_\Sigma,{\cal H}_\Sigma\}_\times =0$.  

An important advantage of Hamiltonian flow and a general property of the Liouville 
equation is:  \\
$\bullet$
G) The normalization and positivity of the probability 
density $\rho$ are conserved in presence of 
a hybrid interaction; hence, its interpretation remains valid. 

\section{Quantum control by a classical time machine}
\label{QuContr}
Our aim here is to combine the results on discrete mechanics (Sect.~\ref{DiscrHamil}), 
where time is one of the dynamical variables and which consequently allows to model 
a particular kind of time machines (Sect.~\ref{2.2}), with those on QM-CL hybrids 
(Sect.~\ref{QuClHybr}). We explore in this framework, how such a classical time machine 
interacts with a quantum object.  

As a concrete example, we consider an {\it oscillator-like time machine coupled to 
a q-bit}. The former is represented by the Hamiltonian function: 
\begin{equation}\label{osctimemach} 
{\cal H}_{\mbox{\scriptsize CL}}(x,p;t):=\frac{\zeta}{2}\cos (\omega t) [p^2+\Omega^2x^2]
\;\;, \end{equation} 
cf. Eq.\,(\ref{effH}), where $\Omega$ denotes the proper oscillator frequency,  
while $\omega$ is the frequency of the change of time direction, cf. Sect.~\ref{2.2}, 
and the dimensionless constant $\zeta$ parametrizes its amplitude. For 
$\Omega\gg\omega$, the oscillator performs many oscillations (circles in 
phase space), before the time direction changes; conversely, for $\Omega\ll\omega$, 
the oscillator moves only little before beginning to trace its trajectory in 
phase space in the opposite direction.  Qualitatively similar behaviour of 
the time machine is expected for other than oscillator potentials. 
  
The q-bit is described, in the oscillator representation, cf. 
Sect.~\ref{OscRepr}, by the Hamiltonian function:  
\begin{equation}\label{qbitH} 
{\cal H}_{\mbox{\scriptsize QM}}(X_1,X_2,P_1,P_2):=\frac{E_0}{2}\sum_{i=1,2}(-1)^i(P_i^{\;2}+X_i^{\;2}) 
\;\;, \end{equation} 
with $E_0$ an energy scale, cf. Eq.\,(\ref{HamiltonianQM2}).    
Wave function normalization, Eq.\,(\ref{oscillnormalization}), is required by: 
\begin{equation}\label{qbitnorm} 
2{\cal C}\equiv X_1^{\;2}+X_2^{\;2}+P_1^{\;2}+P_2^{\;2}\stackrel{!}{=}2 
\;\;. \end{equation}

The model is completed by choosing a hybrid interaction, for example: 
\begin{equation}\label{qbitosctime} 
{\cal I}(x,;X_i,P_i) :=\lambda x\cos (\omega t)\langle\Psi |\hat O|\Psi\rangle  
=\lambda x\cos (\omega t)\sum_{i,j=1,2}O_{ij}(X_i-iP_i)(X_j+iP_j)
\;\;, \end{equation} 
using the oscillator expansion of a generic state $|\Psi\rangle$;  
$O_{ij}:=\langle\phi_i|\hat O|\phi_j\rangle$ denotes a matrix 
element of the q-bit observable $\hat O$ ($=\hat O^\dagger$) in the basis of 
energy eigenstates 
corresponding to ${\cal H}_{\mbox{\scriptsize QM}}$ and $\lambda$ is a coupling 
constant. Naturally, other and more general interactions may be considered. 

Then, the following Hamilton equations are obtained in the usual way 
from the hybrid Hamiltonian 
${\cal H}_\Sigma:={\cal H}_{\mbox{\scriptsize CL}}+{\cal H}_{\mbox{\scriptsize QM}}
+{\cal I}$: 
\begin{eqnarray}\label{xeq} 
\dot x&=&p\cos (\omega t) 
\;\;, \\ \label{peq} 
\dot p&=&-\Big (\Omega^2x+\lambda\langle\Psi |\hat O|\Psi\rangle\Big )\cos (\omega t)
\;\;, \\ \label{Xeq} 
\dot X_i&=&(-1)^iE_0P_i
+\lambda x\partial_{P_i}\langle\Psi |\hat O|\Psi\rangle\cos (\omega t) 
\;\;, \\ \label{Peq}
\dot P_i&=&-(-1)^iE_0X_i
-\lambda x\partial_{X_i}\langle\Psi |\hat O|\Psi\rangle\cos (\omega t)
\;\;, \end{eqnarray} 
where we set $\zeta\equiv 1$, which can always be implemented by rescaling time, 
$E_0$, and $\lambda$. 
In agreement with the general result in Eqs.\,(27)--(28) of Ref.\,\cite{me11}, the 
constraint of 
Eq.\,(\ref{qbitnorm}) is conserved under this Hamiltonian flow, 
$\mbox{d}{\cal C}/\mbox{d}t=\{  {\cal C},{\cal H}_\Sigma\}_\times =0$, and, therefore, 
it is sufficient to impose the constraint on the initial conditions of the equations of 
motion (\ref{xeq})--(\ref{Peq}). 

In order to uncover some characteristic features of this hybrid model,  we 
introduce the {\it internal time} variable $\tau (t):=\omega^{-1}\sin (\omega t)$ into  
Eqs.\,(\ref{xeq})--(\ref{peq}). The resulting 
second order equation (for $x(\tau$)) of a driven harmonic oscillator 
can be solved with the help of its retarded Green's function: 
\begin{equation}\label{xsol} 
x(t)=x_1\cos [\Omega\tau (t)+\phi ]
-\lambda\Omega^{-1}\int_{-\infty}^{\tau (t)}\mbox{d}s\;
\sin [\Omega(\tau (t)-s)]\tilde O(s)
\;\;, \end{equation}  
where the first term solves the homogeneous equation, incorporating  
integration constants $x_1$ and $\phi$, and where the inhomogeneity   
is given by: 
\begin{equation}\label{Otilde} 
\tilde O(s):=(X_1P_2-X_2P_1)_{t(s)}
\;\;, \end{equation} 
with $X$'s and $P$'s evaluated at $t(s)$, determined (modulo $\pi /\omega$) by 
$t=\omega^{-1}\arcsin (\omega s)$ ; for simplicity,  
the q-bit obervable has been assumed to be proportional to the spin-1/2 Pauli 
matrix $\sigma_2$, such that $-O_{12}=O_{21}=i/2$ and $O_{11}=O_{22}=0$.  
Using solution (\ref{xsol}) and 
Eq.\,(\ref{xeq}), we obtain: 
\begin{equation}\label{psol} 
p(t)=\mbox{d}x/\mbox{d}\tau =
-x_1\Omega\sin [\Omega\tau (t)+\phi ]
-\lambda\int_{-\infty}^{\tau (t)}\mbox{d}s\;
\cos [\Omega(\tau (t)-s)]\tilde O(s)
\;\;. \end{equation}  

We see explicitly that the time machine travels periodically forwards and backwards 
in time, due to the periodicity of its {\it internal time} $\tau$ 
with respect to the {\it external time} $t$ governing the 
chronology respecting q-bit (described by the $X,P$-variables). Most notably,  
$\dot x$ and $p$ are not always aligned, {\it i.e.}, of same sign. 

However, 
since from one period of forward (or backward) evolution to the next the external time 
increases by $2\pi /\omega$, generally, the value of the function $\tilde O$ in 
Eqs.\,(\ref{xsol})--(\ref{psol}) 
will change accordingly. This implies that the classical time machine that interacts with the 
q-bit, will go backwards in $(x,p)$ phase space in a different way than it came!  
Which can entail known paradoxes of time travel, such as the {\it grandfather paradox} 
or the {\it unproved theorem paradox} \cite{Lloydetal11a,Lloydetal11b}.   

Summarizing, the interaction of a classical 
time machine with a chronology respecting system, the q-bit here, introduces an 
aspect of ``ageing'' into its dynamics: despite going forwards and backwards in 
time, in general, its state does evolve and depends on the external time $t$. 

Unlike solutions to such paradoxes proposed in the literature for 
{\it quantum systems} in the presence of closed timelike curves (CTCs) by 
Deutsch \cite{Deutsch91}, Lloyd and collaborators 
in the form of post-selected teleportation (P-CTCs) \cite{Lloydetal11a,Lloydetal11b}, 
or the consideration of open timelike curves  (OTCs) by Ralph and collaborators 
\cite{Ralph12}, our model of a {\it classical} time machine does 
not provide enough freedom to eliminate paradoxical situations 
(Novikov principle) by imposing additional constraints on its 
dynamics. Apparently, {\it quantum-classical hybrids} do not work here. 
-- Considering a suitably constrained {\it ensemble} of 
classical time machines might help. However, its physical 
relevance remains to be seen.  

Of course, 
having a QM time machine consistently interacting with a classical object is not ruled 
out by the present model. In fact, previously considered  CTC scenarios should reduce 
to such a hybrid situation under suitable circumstances. 

We consider the effect of the time machine on the q-bit next.  
In this case,  we conveniently  rewrite 
Eqs.\,(\ref{Xeq})--(\ref{Peq}) by undoing the oscillator expansion, 
cf. Eq.\,(\ref{oscillexp}): 
\begin{equation}\label{s12}
i\frac{\mbox{d}}{\mbox{d}t}
\left (\begin{array}{c} 
X_1+iP_1 \\ 
X_2+iP_2 \end{array}\right ) 
=\left (-E_0\hat\sigma_3+\lambda x(t)\cos (\omega t)\hat\sigma_2\right )
\left (\begin{array}{c} 
X_1+iP_1 \\ 
X_2+iP_2 \end{array}\right ) 
\;\;, \end{equation}
where $\hat\sigma_{2,3}$ are the imaginary and diagonal spin-1/2 Pauli matrices, 
respectively. This is a Schr\"odinger equation  representing a 
spin-1/2 in a {\it magnetic field}. In particular, here its 2-component is time dependent.  
With the time dependence arising from $x(t)\cos (\omega t)$, 
cf. Eqs.\,(\ref{xsol})--(\ref{Otilde}),  this {\it effective  
Schr\"odinger equation is nonlinear and non-Markovian}.   

The nonlinear and non-Markovian behaviour can be neglected for sufficiently 
small coupling $\lambda$, in which case the time dependence of the effective 
magnetic field is given by the following factor: 
\begin{equation}\label{B}  
B(t):=\lambda x_1\cos (\omega t)\cos\left (\Omega\omega^{-1}\sin (\omega t)\right ) 
\;\;, \end{equation} 
incorporating only the first term from the right-hand side of Eq.\,(\ref{xsol}).  
Concerning the q-bit, the corresponding instantaneous eigenvalues of the effective 
Hamiltonian are shifted in this approximation (lowest nonvanishing order 
in $\lambda$) and are simply given by: 
\begin{equation}\label{energyeigen}
E_\pm =\pm E_0\left (1+B^2(t)/2E_0^{\;2}\right )
\;\;. \end{equation}   
This result would, in principle, allow to constrain parameters defining the present 
toy model of a quantum-classical hybrid, consisting of a classical time machine 
interacting with a q-bit, given the manifold laboratory realizations of q-bits. 

Furthermore, among others, there are generalizations of the  hybrid interaction, 
Eq.\,(\ref{qbitosctime}), which could give rise to a {\it rotating magnetic 
field} instead of the oscillating one in Eqs.\,(\ref{s12})--(\ref{B}). This, in turn, produces 
effects like a {\it Berry phase}, or its generalizations (see, {\it e.g.}, the recent 
Ref.\,\cite{BerryGen12} and references therein), which could 
serve as well to constrain such models.  

However, as we have discussed, the classical time machines addressed here are 
likely bound to reproduce the paradoxes of time travel. If they are not directly observable, for some reason, their indirect effects on quantum systems may be 
worth further  study.   

\section{Conclusions}
\label{Conclu} 
Our purpose here has been to explore the possibility that classical ``time machines''  
couple directly to quantum mechanical objects. 

We invoked the discrete mechanics proposed by T.D. Lee, in which time belongs to 
the set of dynamical variables \cite{Lee83,Lee87}. Suitably modifying the underlying 
action, we have developed a Hamiltonian theory of such {\it discrete classical dynamical 
systems}. 
Choosing the dynamics of the time variable appropriately, we are led to systems which 
evolve forward and backward in time, {\it time machines} or, more precisely, 
time reversing machines. 

In the 
continuum limit and for particular choices of the dynamics of time, the motion is 
periodic. Thus, such an object evolves forward in time, forming a trajectory in phase 
space, until it comes to a halt, then traces this trajectory backwards in time, comes to 
halt, evolves forward again, and so on. A clock carried on board would be seen running 
alternatingly forwards and backwards.  -- These time machines are distinct from 
the closed timelike curves (CTCs) on which an object travels, which have been 
frequently discussed in the literature, see Refs. \cite{Lloydetal11a,Lloydetal11b}, 
for example, and works referred to there. They might be realizable in physical analogue 
models.  
  
In order to describe the direct coupling of such a time machine with a quantum object, 
{\it e.g.} a q-bit, we reviewed our recent proposal for a consistent quantum-classical hybrid
 dynamics, which is based on a phase space formalism for classical as well as for 
quantum mechanics \cite{me11,me12,me12b,Heslot85,Buric12}. 

We have defined a toy model of such a QM-CL hybrid, consisting of 
an oscillator like classical time machine coupled to a q-bit and discussed its 
Hamiltonian equations of motion. While this could lead to observe the action of a 
time reversing machine through its effects on a quantum object, we have also 
pointed out that common time travel paradoxes would affect  
the classical time machine. 

In retrospect, the latter is understandable, since 
the outcome of the evolution of a classical object is deterministic and fixed, 
{\it e.g.}, by initial conditions, 
to the extend that no additional (nonlinear) constraints can be imposed, as in 
the quantum case. The QM-CL hybrids that we described do {\it not} 
alter this circumstance. For quantum mechanical objects travelling on CTCs, however, such 
constraints serve to suppress the paradoxes by selecting well-behaved ones 
from the ensemble of all possible histories \cite{Lloydetal11a,Lloydetal11b,Deutsch91}. 
Which poses the question whether a {\it quantum mechanical time reversing machine}, 
based on the Hamiltonian discrete meachanics presented here,  
can similarly avoid time travel paradoxes?  

\vskip 0.5cm \noindent 
{\bf Acknowledgements:} It is a pleasure to thank M. Crosta, M. Gramegna and M. 
Ruggiero for the invitation to give a talk in the inspiring atmosphere of the conference 
{\it Time Machine Factory} (Torino, October 2012) and to thank N. Buri\'c, L. Maccone,  
and C. Stoica for discussions and correspondence.

\end{document}